\documentclass[conference]{IEEEtran}
\ifCLASSINFOpdf
   \usepackage[pdftex]{graphicx}
\else
\fi
\begin{document}
%
\title{Software Engineers' Information Seeking Behavior in Change Impact Analysis -- An Interview Study}


\author{\IEEEauthorblockN{Markus Borg}
\IEEEauthorblockA{RISE SICS AB\\
Lund, Sweden\\
markus.borg@sics.se}
\and
\IEEEauthorblockN{Emil Al\'egroth}
\IEEEauthorblockA{Blekinge Institute of Technology\\
Karlskrona, Sweden\\
emil.alegroth@bth.se}
\and
\IEEEauthorblockN{Per Runeson}
\IEEEauthorblockA{Lund University\\
Lund, Sweden\\
per.runeson@cs.lth.se}
}


%


\maketitle

\begin{abstract}
Software engineers working in large projects must navigate complex information landscapes.
Change Impact Analysis (CIA) is a task that relies on engineers' successful information seeking in databases storing, e.g., source code, requirements, design descriptions, and test case specifications.
Several previous approaches to support information seeking are task-specific, thus understanding engineers' seeking behavior in specific tasks is fundamental.
We present an industrial case study on how engineers seek information in CIA, with a particular focus on traceability and development artifacts that are not source code.
We show that engineers have different information seeking behavior, and that some do not consider traceability particularly useful when conducting CIA.
Furthermore, we observe a tendency for engineers to prefer less rigid types of support rather than formal approaches, i.e., engineers value support that allows flexibility in how to practically conduct CIA.
Finally, due to diverse information seeking behavior, we argue that future CIA support should embrace individual preferences to identify change impact by empowering several seeking alternatives, including searching, browsing, and tracing.
\end{abstract}


%
\IEEEpeerreviewmaketitle

\section{Introduction}
\label{sec:intro}

The information landscape of a large Software Engineering (SE) project is complex. 
First, the \textit{sheer volume of information} that developers maintain in large projects threatens the overview, as tens of thousands of development artifacts are often involved. 
Consequently, the scale of the information landscape in large SE projects typically exceeds the individual developer's comprehension~\cite{robillard_introduction_2014}.

Second, developers work collaboratively on \textit{heterogeneous development artifacts} stored in various software repositories such as source code repositories, requirements databases, test management systems, and general document management systems.
Often the databases have \textit{poor interoperability}~\cite{estublier_software_2000}, thus they turn into ``\textit{information silos}'', i.e., simple data storage units with little transparency for other tools.

Third, as source code is easy to modify, at least when compared to accompanied hardware, the software system under development continuously evolves during a project. 
Not only does the source code evolve, but the related development artifacts should also co-evolve to reflect the changes, e.g., design documents and test case descriptions might require continuous updates~\cite{cleland-huang_event-based_2003}.
Large software systems might evolve for decades
introducing both versioning problems and obsolete information.
Consequently, staying on top of the information landscape in large SE projects constitutes a significant challenge for both developers and managers~\cite{robillard_introduction_2014}.

This is particularly challenging in the domain of safety-critical systems, where mandatory practices, such as \textit{Change Impact Analysis} (CIA) rely on navigating the information landscape.
If the project environment does not provide sufficient support for navigation and retrieval, considerable effort is wasted on locating the relevant information~\cite{karr-wisniewski_when_2010}.
Unfortunately, large SE projects are threatened by information overload, i.e., ``\textit{a state where individuals do not have the time or capacity to process all available information}''~\cite{eppler_concept_2004}.
Consequently, an important characteristic of a software project is the \textit{findability} it provides, i.e., ``\textit{the degree to which a system or environment supports navigation and retrieval}''~\cite{morville_ambient_2005}.

Poor findability has several negative consequences in SE projects.
Freund \textit{et al.} reported that software engineers spend about 20-30\% of their time consulting various software repositories, but still often fail to fulfil their information needs~\cite{freund_modeling_2005}. 
Dagenais \textit{et al.} showed that poor search functionality in information repositories constitutes an obstacle for newcomers entering new software projects~\cite{dagenais_moving_2010}.
Bjarnason \textit{et al.} discovered that the sheer volume of information threatens the alignment between requirements engineering and testing in large projects~\cite{bjarnason_challenges_2014}.

In this paper, we focus on CIA of development artifacts that are \textit{not} source code. 
We conduct a case study with two units of analysis on software engineers' information seeking in CIA.
This contribution is part of an ongoing series of works to support CIA in a safety-critical context, thus it should be interpreted along with our previous publications\footnote{We share the same interviewees and use the same IDs as in~\cite{borg_supporting_2016}: A-N}~\cite{borg_practitioners_2016,borg_supporting_2016}.
We show that engineers seek information in several ways, including searching and browsing databases, asking colleagues, and consulting formal trace links.
With a particular forcus on traceability and CIA, we report that engineers perceive the usefulness of trace links differently.
Our study identifies two concrete improvement areas for the case company: 1) improving the issue tracker, and 2) developing information policies for the document management systems.
Finally, our empirical findings contribute to the understanding of engineers' information seeking in a specific task, thus supporting development of task-specific search solutions~\cite{freund_modeling_2005,grzywaczewski_task-specific_2012}.

The rest of this paper is organized as follows.
Section~\ref{sec:bw} introduces CIA, traceability, and related work on information seeking in SE.
Section~\ref{sec:case} presents CIA in the context of the case company.
Section~\ref{sec:method} describes the research method, and Section~\ref{sec:res} presents our findings.
Finally, Section~\ref{sec:threats} shares the main threats to validity, and Section~\ref{sec:conc} concludes the paper and outlines future work.

\section{Background and Related Work}
\label{sec:bw}
This section reports the fundamentals of CIA and discusses how the literature proposes traceability as a key solution.
We also present related work on information seeking in SE.

\subsection{Change Impact Analysis and Traceability}
A popular definition of CIA is provided by Bohner, stating that CIA is ``\textit{identifying the potential consequences of a change, or estimating what needs to be modified to accomplish a change}''~\cite{bohner_software_1996}.
CIA can be described as a cognitive process of incrementally adding items to a set of candidate impact, which is known to be both tedious and error-prone for large systems~\cite{mens_introduction_2008,lehnert_taxonomy_2011}. 
However, as CIA is mandated by most safety standards, companies aspiring to release certified software systems must comply.  

Most CIA work in industry is manual~\cite{de_la_vara_industrial_2016}, although the importance of improved CIA tools has been highlighted in research for a long time~\cite{bohner_software_2002}. 
Also, two recent reviews of scientific literature shows that most research on CIA is limited to impact on source code~\cite{lehnert_taxonomy_2011,li_survey_2013}.
However, as stated in Lehnert's review: ``\textit{more attention should be paid on linking requirements, architectures, and code to enable comprehensive CIA}''~\cite[pp. 26]{lehnert_taxonomy_2011}.
Especially in safety-critical development, it is critical to also analyze how a change to a software system affects artifact types that are not source code, e.g., whether any requirements are affected, or which test cases should be selected for regression testing.

\noindent \textit{Traceability} has been discussed in SE since the pioneering NATO Working Conference on SE in 1968.
Randall argued that a developed software system should ``\textit{contain explicit traces of the design process}''~\cite{randall_towards_1969}.
Boehm mentioned traceability as an important contemporary SE challenge in a state-of-the-art survey from 1976, and predicted traceability to become a future research trend~\cite{boehm_software_1976}.
Concurrently, industrial practice acknowledged traceability as a vital part of high-quality software, and by the 1980s several development standards had emerged that mandated traceability maintenance~\cite{dorfman_standards_1994}.

In the 2000s, the growing interest in agile development methods made many organizations downplay traceability.
Agile developers often consider traceability management to be a burdensome activity that does not generate return on investment~\cite{cleland-huang_traceability_2012}.
Still, traceability remains non-negotiable in development of safety-critical systems.
Safety standards such as ISO 26262 in the automotive industry\footnote{ISO 26262-1:2011 Road vehicles -- Functional safety} and IEC 61511 in the process industry sector\footnote{IEC 61511-1 ed. 1.0 Safety Instrumented Systems for the Process Industry Sector} explicitly requires traceability through the development lifecycle.

While traceability management is costly, several studies show that access to trace links support developers' CIA.
Access to traces can support developers with CIA~\cite{li_requirement-centric_2008,cuddeback_automated_2010}, which is often often used to motivate mostly traceability efforts within SE projects. 
De Lucia \textit{et al.} state that trace links help engineers to understand relationships and dependencies among development artifacts, and that when a feature needs to be changed ``traceability helps to locate the pieces of design, code and whatever needs to be maintained''~\cite[pp. 1]{de_lucia_traceability_2008}.
Cleland-Huang \textit{et al.} proposed a solution for event-based traceability, connecting development artifacts through publish-subscribe relationships~\cite{cleland-huang_event-based_2003}, i.e., when an artifact is changed, the subscribers are informed.
Event-based traceability is an example of work on how to bring practical value to stored trace links, another example is our previous work on ANONYMIZED FOR REVIEW~\cite{anonymous_details_2016}

\subsection{Information Seeking in Software Engineering}
As SE work is knowledge-intensive, quick and concise access to information is essential.
Although information seeking\footnote{Also referred to as: information access, findability, and enterprise search.} has been the target of previous research in SE, most work on engineers' information seeking behavior relate to engineering disciplines in general, e.g., the comprehensive review by King \textit{et al.}~\cite{king_communication_1994}.
One observable behavior is the ``principle of least effort'', i.e., that engineers prefer oral/internal communication over written/external, typically explained by faster access.
Somewhat more recently, Hertzum and Pejtersen conducted a multiple case study on information seeking in two engineering organizations: Novo Nordisk and Danfoss~\cite{hertzum_information-seeking_2000}.
They found that engineers intertwine looking for informing documents and looking for informed people;
One often leads to the other.
Also, they found that engineers often interact socially to get information without engaging in explicit searches.
They conclude that not only document retrieval must be supported, but also searching for people is important. 

SE has some unique characteristics compared to general engineering disciplines, thus several researchers have targeted information seeking behavior of software engineers.
Hertzum did an observational study encompassing 16 meetings and in total 580 instances of information seeking~\cite{hertzum_importance_2002}.
He argues that ``principle of least effort'' does not apply to software engineers because of quick information access;
Instead, the main reason is that software engineers have more trust in close-by, internal information sources.
Freund \textit{et al.} conducted an empirical study of software consultants to understand how contextual factors shape the information seeking behavior of software engineers~\cite{freund_modeling_2005}.
They found that that how engineers' seek information is shaped by the variety of document types and information channels, and presented a four-level model of their work context, which they consider a prerequisite for developing task specific search solutions.
Grzywaczewski and Iqbal also did work on task specific search systems~\cite{grzywaczewski_task-specific_2012}, stressing the need to first perform detailed analyses of how software engineers seek information.
Milewski conducted a survey of 84 software engineers to explore cultural differences regarding information seeking between the US, Europe, East Asia, and India/Pakistan~\cite{milewski_global_2007}.
He found a general tendency for software engineers to favor reading documentation for tasks where the goal is to seek factual information, but asking colleagues is preferred when seeking information to diagnostic tasks.
Furthermore, Milewski reports to differences: 1) engineers in the India/Pakistan group were more inclined to ask colleagues for factual tasks, and 2) Europeans more often study documentation for diagnostic tasks.

Several studies have targeted information seeking in software maintenance.
One example is the six-step model of developers' information seeking by Buckley \textit{et al.}~\cite{buckley_empirically_2006}.
Their model suggests that developers seek information in a highly iterative fashion, both skipping steps and returning to previous steps, as they adapt to the task at hand.
In an observational study of 17 developers over 25 hours, Ko \textit{et al.} identified 334 instances of information seeking and 21 different information needs~\cite{ko_information_2007}.
``Submitting a Change'' is one of seven clusters of information needs reported, but it does not cover CIA.
Ko \textit{et al.} also studied how developers seek information when correcting issues~\cite{ko_exploratory_2006}.
However, neither this study addresses CIA, and in contrast to our work the focus is on navigating source code.
While Ko \textit{et al.}'s work is primarily on the source code level, their early application of Information Foraging Theory (IFT) in SE created a foundation for later pieces of related work.

IFT explains and predicts how people navigate in response to the information in their environment~\cite{fleming_information_2013}.
The theory posits that users adapt their search strategies to maximize gains of valuable information per time unit.
The central mechanism in IFT is information scent: the perception of value  of information sources based on ``proximal cues'', e.g., a code comment, a hyperlink, or a document ID.
Lawrence and colleagues have repeatedly used IFT in software maintenance, e.g., to successfully predict how developers navigate source code when debugging~\cite{lawrance_how_2013}.
Again, the research is focused on source code, and they explicitly state non-code artifacts as a direction for their future work.
Niu \textit{et al.} also explored IFT in software maintenance, more specifically in vetting of traceability matrices containing trace links between requirements and source code~\cite{niu_departures_2013}, rather similar to our focus on CIA.
They studied two foundational optimality models in IFT: 1) making optimal decisions on what information to consume and what to ignore, and 2) finding the optimal time to seek information.
From our perspective, Niu \textit{et al.}'s most important finding is that the six student subjects in their experiment explored more information scents than what the optimal model suggests, and that they revisited the same scent several times.


A number of studies on SE information seeking has targeted issue reports, closely related to our study as the CIA process in the case company is tightly connected with the issue tracker (see Section~\ref{sec:case}).
The number of incoming issue reports in large software engineering projects can be overwhelming, and particularly research on duplicate detection has received much attention.
Runeson \textit{et al.} pioneered issue duplicate detection using standard information retrieval techniques~\cite{runeson_detection_2007}.
Several researchers have done similar work, including a replication by Borg \textit{et al.} that discusses improving findability in the issue tracker using the open source software search library Apache Lucene~\cite{borg_replicated_2014}.
Also in issue tracking there is research available on task specific search solutions, e.g., Baysal \textit{et al.} developed customized issue dashboards based on a grounded theory study of developers' comments~\cite{baysal_no_2014}.

As presented in this section, tool support for information seeking is often highly task-dependent, which is also emphasized in research on recommendation systems in SE~\cite{robillard_introduction_2014}. Hence there is a need for studies that investigate specific tasks.
Our study contributes by specifically investigating such an important task: software engineers' non-code information seeking in CIA.

\section{Case Description}
\label{sec:case}

The case company develops safety-critical industrial automation systems. 
The system under study has its roots in the 1980s and fulfil the IEC 61511 standard via Safety Integrity Level (SIL) 2 certification, according to the IEC 61508 standard\footnote{IEC 61508 ed 1.0, Electrical/Electronic/Programmable Electronic Safety-Related Systems}.
The developed software system must be of high quality, therefore all changes to source code have to be analyzed before committing. 
Moreover, detailed system documentation is maintained, mapped to the vertical abstraction layers in the V-model. 
The projects follow a rigid development process with hundreds of collaborating engineers distributed globally, while the two main development sites are in Sweden and India. 
The software system contains over one million lines of code, dominated by C/C++ and some extensions in C\# and VB. 

Prioritized features originating from various customers (and sometimes pre-ordered feature requests) are incrementally added and extensively tested. 
When developing new features, and when fixing issues related to existing features, several changes are made. 
When the development is finished, the development organization needs to present a safety case for an external assessor, demonstrating that the system is acceptably safe for a given application in a given operating environment. 
The set of documented CIA analyses is a crucial component of the \textit{safety case}. 
Therefore, the safety engineers at the case company have developed CIA report template (cf. Table~\ref{tab:ciatemplate}), containing questions that must be answered for each CIA, to support the safety case in relation to the IEC 61508 safety certification. 
The developers use this template to document their CIA before committing any source code changes.

\begin{table}
\centering
\caption{The CIA template used in the case company, first presented by Klevin~\cite{klevin_people_2012}.}
\begin{tabular}{|l|p{7.5cm}|}
\hline			
\multicolumn{2}{|c|}{Change Impact Analysis Questions}		\\
\hline	
1	&	Is the reported problem safety-critical?	\\
\hline			
2	&	In which versions/revisions does this problem exist?	\\
\hline			
3	&	How are general system functions and properties affected by the change?	\\
\hline			
4	&	List modified code files/modules and their SIL classifications, and/or affected safety safety related hardware modules.	\\
\hline			
5	&	Which library items are affected by the change? (e.g., library types, firmware functions, HW types, HW libraries)	\\
\hline			
6	&	Which documents need to be modified? (e.g., product requirements specifications, architecture, functional requirements specifications, design descriptions, schematics, functional test descriptions, design test descriptions)	\\
\hline			
7	&	Which test cases need to be executed? (e.g., design tests, functional tests, sequence tests, environmental/EMC tests, FPGA simulations)	\\
\hline			
8	&	Which user documents, including online help, need to be modified?	\\
\hline			
9	&	How long will it take to correct the problem, and verify the correction?	\\
\hline			
10	&	What is the root cause of this problem?	\\
\hline			
11	&	How could this problem been avoided?	\\
\hline			
12	&	Which requirements and functions need to be retested by the product test/system test organization?	\\
\hline			
\end{tabular}		
\label{tab:ciatemplate}														
\end{table}		

The CIA process is tightly connected with the issue management process, as all changes to formal development artifacts require an issue report in the issue repository. 
All completed CIA reports are stored in the issue tracker as attachments to issue reports. 
Developers typically access the issue tracker using a simple web interface.
Other important components in the information landscape are the Document Management System (DMS), the source code repository (also containing test code), and the intranet

\section{Method}
\label{sec:method}
Four research questions quide this study, all of them studied in the context of identifying change impact as part of corrective software maintenance, i.e., resolving open issue reports:

\begin{itemize}
\item RQ1. How is traceability used in CIA?
\item RQ2. What CIA support is available in the case company?
\item RQ3. What information seeking behaviors exist in CIA?
\item RQ4. How do engineers conclude a CIA as satisfactory?
\end{itemize}

\begin{figure}
\centering
\includegraphics[width=0.5\textwidth]{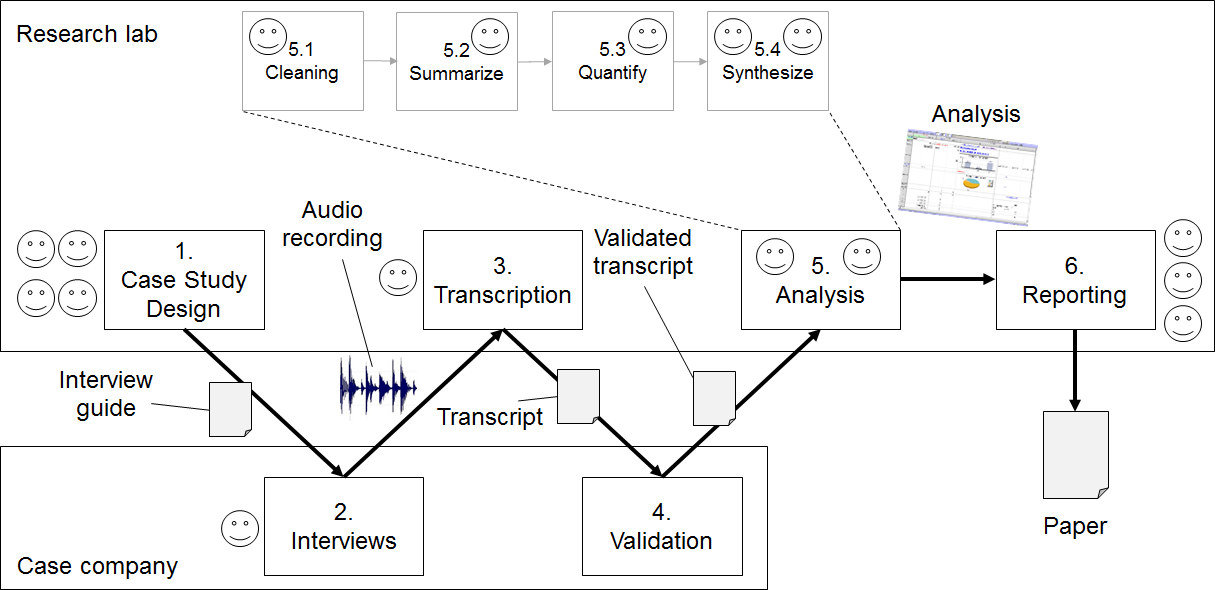}
\caption{Overview of the study. Smileys depict the number of researchers involved in each step.}
\label{fig:overview}
\end{figure}

We conducted a multiple unit industrial case study since the studied phenomenon could not be separated from its context~\cite{runeson_case_2012}.
The case under study is the CIA activity, and two development teams constitute the units of analysis: Unit Sweden and Unit India.
In addition, we included three senior engineers that do not belong to any teams, referred to as Senior Experts.
Figure~\ref{fig:overview} shows an overview of the study.
Four researchers iteratively (step 1) designed the case study and documented it in a case study protocol.
We created an interview guide\footnote{http://serg.cs.lth.se/fileadmin/serg/ImpRec\_EvalStudy/} for semi-structured interviews to be able to ask both closed and open-ended questions. 
We asked open questions in the beginning and end of the interviews, according to the time glass interview model~\cite{runeson_case_2012}.
Now we analyze a subset of the questions asked: Pre2 e)-g) in the interview guide.

The data collection (step 2) consisted of interviews in Swedish or English, for confidentiality reasons conducted by a single researcher. 
The same researcher transcribed all interviews word by word (step 3) and sent them back to the interviewees for validation (step 4).
We interviewed 14 engineers, referred to as `Int A-N', of whom 10 are developers that work with source code.
More specifically, in Sweden we interviewed: one R\&D manager, one safety engineer, three senior developers (incl. the team leader), and one junior developer, and in India: one product manager, one technical manager, four senior developers (incl. the team leader), and two junior developers.
As a preliminary analysis (step 5), the transcripts were copied into a spreadsheet structured according to the interview guide, and longer answers were divided into smaller chunks. 
The spreadsheet was then cleaned (step 5.1) to remove redundancy of spoken language, as well as non-informative pieces\footnote{Examples include: `eh', `like', and `you know...'.}.

To mitigate researcher bias, the cleaned chunks were analyzed using an exploratory analysis approach by an independent researcher, i.e., a researcher that had not been part of the previous steps. 
The analysis was performed in three substeps for each question of the survey.
First, the cleaned chunks for each question were analyzed individually and summarized (step 5.2) in a compact form. 
In this substep, the independent researcher extracted the core perception of the Int, e.g., the Int appeared knowledgeable about how traceability was achieved in the project, and that a positive perception was communicated.
The independent researcher then quantified (step 5.3) the extracted perception on a scale between 1-10 to provide a nuanced view. 
In the final analysis substep, the results were synthesized (step 5.4) to draw conclusions based on the qualitative data;
First by the independent researcher, then validated by the first author.
Finally, we report our results in this publication (step 6).

The qualitative analysis is inspired by grounded theory as conclusions are compiled from the raw data. 
In contrast to traditional grounded theory, however, no traditional coding was used;
The raw data were already organized by the interview design, i.e., naturally grouped or `coded' by the interview questions.
As is essential in qualitative analysis, we have a clear chain of evidence from conclusions to RQs down to individual quotes given by the Ints.


\section{Results and Discussion}
\label{sec:res}
In this section we present our findings, organized per RQ.

\subsection{Traceability and CIA (RQ1)}
\label{sec:RQ1}
As the importance of traceability to CIA has been established in several studies, we addressed it first in the interviews; 
Our goal was to collect the Ints' thoughts on the value of traceability in CIA before influencing them with the rest of our interview questions.
As part of RQ1, we also determine what types of trace links the engineers use in CIA. 

\begin{table}
\centering
\caption{Overview of RQ1. Black boxes show ``very useful'', striped boxes ``moderately useful'', and white boxes show ``not useful''.}
\includegraphics[width=0.3\textwidth]{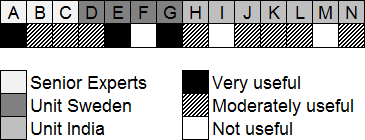}
\label{tab:rq1}
\end{table}

Table~\ref{tab:rq1} shows the perceived usefulness of traceability in CIA. 
Three out of 14 Ints (A, E, G) regard traceability as very useful.
Int A praises the considerable efforts spent on improving traceability in recent years (triggered by the second edition of IEC 61508), including bidirectional trace links often maintained in a new architecture modelling tool: ``when I started here 10 years ago we didn't have a proper architecture specification. Now we have two-way traces across all levels in the V-model... acquired through much effort''.
He explains that the traceability information is complete, but requests better tools for a subset of the traces, i.e., from specific source code files to test specifications;
those links are currently maintained manually in a large spreadsheet (hereafter called the `T-sheet').
Ints E and G both report that they trace effectively using documents; specific tables (hereafter called `T-tables') in Design Descriptions (DD) refers to Detailed Design Descriptions (DDD), which in turn contains trace links to individual requirements. 
Finally, the individual requirements are mapped to the test specifications that verify them.
While Int E thinks the trace links are easy to find in the documents, Int G prefers using the new architecture modelling tool.

Three Ints (F, I, M) state that traceability does not support their CIA work.
Int F explicitly complains that traces are mostly inaccessible, thus not helpful for developers.
While he supports the new architecture modelling tool, he is very critical about storing trace links in individual documents, i.e., the traditional way of storing trace links in the case company, still used for parts of the system.
Furthermore, Int F points out that the tracing is obstructed by the missing interoperability between the architecture modelling tool and the DMS: ``We had a better overview at my previous company, you searched for documents directly in the modelling tool. Here you instead find a doc-ID, search for it in the DMS... Find other references, open a second document and then a third''.
Int M prefers to explore how previous issues were resolved rather than investigating formally maintained traces. 

Most Ints consider traceability somewhat helpful in CIA, or do not have a strong opinion at all.
Ints C and H use both trace links stored in documents and a traceability matrix (hereafter called the `T-matrix') maintained for a specific part of the product managed in Unit India. 
However, they have contrasting views on the T-matrix' trustworthiness: Int H trusts the trace links, but Int C often accepts them although he questions their completeness.
Int J also uses the T-matrix in CIA, but mainly to report which test cases to execute to verify changes.
Int D prefers to trace on a feature level rather than the more detailed requirements level, although most traceability effort resides on maintaining the fine-granular traces links.
He mainly follows traces stored in the architecture model, and only reluctantly looks for traces in documents.
On the contrary, Int K reads the T-tables in DDDs to identify trace links.
Ints L and N do not value the formally maintained trace links much in their CIA, as they primarily work will small components, i.e.,  staying on top of their development and its documentation is not a problem: ``the [anonymized component's document space] is not that big. The requirements are very compact and to the point, so it's not difficult to navigate'' (Int L).
Finally, Int B, representing a management perspective, argues that the latest revision of IEC 61508 has made traceability more useful for CIA due to the introduction of bidirectional traces.
On the other hand, he explains that the costs of traceability maintenance have increased accordingly, but it is simply a price they must pay to comply with the IEC 61508 standard.

\begin{figure}
\centering
\includegraphics[width=0.5\textwidth]{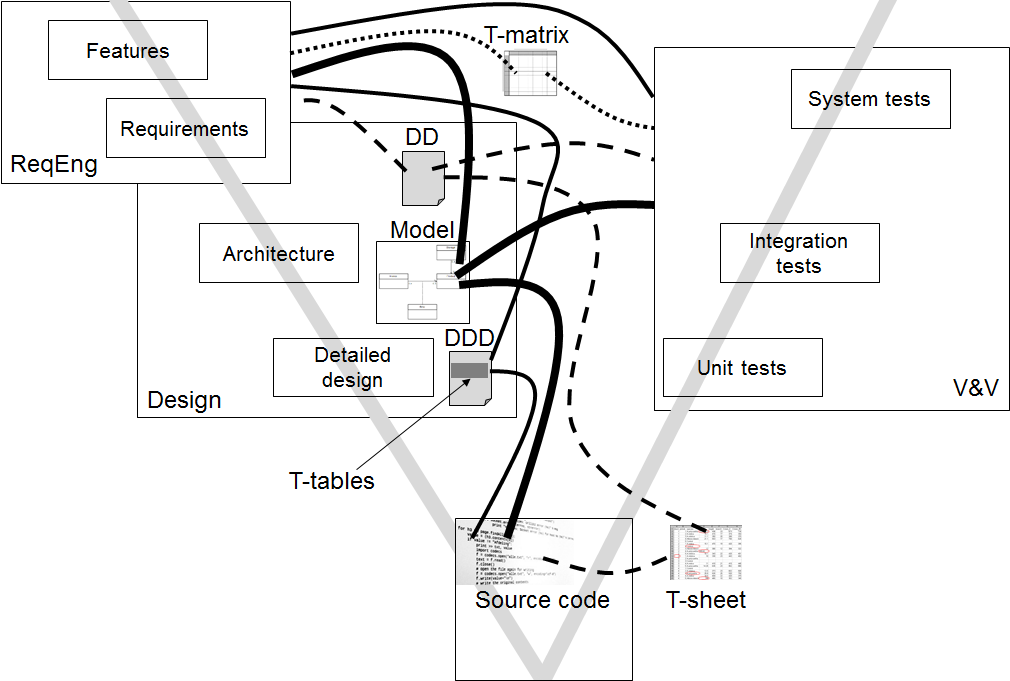}
\caption{Overview of trace links used in CIA. Bold line: code-model-ReqEng/V\&V (Ints A, D, G). Dashed line: code-T-sheet-DD-ReqEng/V\&V (Ints A, E). Solid line: code-DDD-ReqEng-V\&V (Ints D, G, H, K). Dotted line: ReqEng-T-matrix-V\&V (Ints C, H, J, K).}
\label{fig:traces}
\end{figure}

As part of RQ1, we also explored more concretely what types of trace links are used by engineers in CIA;
Figure~\ref{fig:traces} depicts an overview of our findings.
According to Int A, the safety engineer, the prescribed process is the following:
Developers start from the source code and navigate to the design level.
For new parts of the system, developers locate the related component in the architecture model (cf. bold line in Figure~\ref{fig:traces}) from which you find trace links to requirements and V\&V.
For older parts of the system, i.e., not covered by the architecture model, developers identify the source code files in the T-sheet specifying the corresponding SIL and its related DDs (cf. dashed line).
From the DDs, the developer finds trace links to requirements and test case specifications (cf. V\&V in Figure~\ref{fig:traces}).
Talking to Int E confirms that the T-sheet is used to find change impact, and Ints D and G report tracing via the architecture model.

Several Ints (D, G, H, K) mention also another trace link path (cf. solid line): from the source code to the DDD.
The DDDs contain a T-table with trace links to the requirements, and all requirements are mapped to test cases that verify them.
Ints in Unit India report that they map requirements to test cases using a fairly new T-matrix for their part of the system (cf. dotted line); it is considered useful but costly to maintain.
Finally, two Ints (B and E) explain that often they already know the trace links by heart, thus they do not need to consult any traceability information at all during the CIA.

Our answer to RQ1 is that engineers use several different types of trace links in CIA (cf. Fig~\ref{fig:traces}).
However, the engineers in the case study do not perceive the trace links as useful for CIA as one would expect (cf. Table~\ref{tab:rq1}), given the widely positive claims stated in previous work~\cite{de_lucia_traceability_2008,li_requirement-centric_2008,cuddeback_automated_2010};
Since the case company has spent considerable effort on traceability in recent years, i.e., highly accurate bidirectional trace links are maintained, we expected the engineers to value them more in CIA.
Thus, our results show that companies must actively work on making stored trace links useful to engineers, simply storing them in documents or databases might not be enough to get return on investment (ROI).

Our study shows that some engineers prefer to conduct CIA without consulting formal trace links.
Especially the interviewees in Unit India express that they can overview their development context also without trace links, i.e.,  they are responsible for their documents, and they know when and how changes impact other artifacts.
As components with fewer dependencies are developed in Unit India, this comes as no surprise: the challenge of CIA is amplified by  increased complexity~\cite{bohner_software_2002}.  
Furthermore, we hypothesize that some developers consider the trace links to be more of a burden than a support in the context of CIA.
Although further research is needed to support the claim, traceability maintenance is known to be tedious~\cite{cleland-huang_event-based_2003,de_lucia_traceability_2008}, and several Ints indicate that they prefer to navigate in candidate change impact based on their experience and system comprehension;
more formal traces in combination with a mandated process to manually investigate them inevitably leads to additional work.
Future research is needed on how to customize cost-effective traceability that is fit-for-purpose for CIA in an organization, resonating with Cleland-Huang \textit{et al.}'s first two goals for future traceability research~\cite{cleland-huang_software_2014}: 1) trace links should support stakeholder needs, and 2) the ROI from using traces should be adequate in relation to the outlay of establishing it.

Regarding the types of trace links used in CIA, we notice that forward tracing to the V\&V side of the V-model dominates;
Although bidirectional traceability is maintained, no Int appears to navigate backwards from test case specifications to design and requirements.
Apparently, tracing backwards from source code to design and requirements, and then forwards to test cases, is the standard approach in CIA.
This is somewhat surprising, as developers in the same case company have explained that they are less knowledgeable about requirements~\cite{anonymous_details_2016}, and Int A explains that developers are primarily skilled at navigating code: ``seeking in the code base... the developers tend to master the related tools well''. 
As they are more comfortable navigating source code and test code, finding requirements from the test side might be a possibility, i.e., from test code to their closely related test specifications and then across the V-model to the requirements side.
Future work should further explore the potential of test-to-requirements tracing in CIA, which represents a fairly unexplored type of backwards tracing within traceability research. 



\subsection{Current CIA Support (RQ2)}
In 12 out of 14 interviews we explicitly asked what kind of CIA support was currently available (cf. Table~\ref{tab:rq2}).
Two of the Ints (A and B) were clearly positive about the current CIA support.
Int A complimented the considerable efforts in recent years to establish traceability from requirements to the design model to the source code;
According to him, the trace links are maintained properly, and he stated: ``our large traceability effort should be useful [in CIA] now, otherwise we have a problem''. 
Int B emphasizes the value of the formal training provided for new employees, including an explicit session on CIA as part of the change management training: ``you must master the tools we use, learn how we model the design, our requirements structure... You must learn why we do CIA, and what they should contain.''
Moreover, Int B explains that the organization supports informal knowledge transfer to new employees from senior colleagues covering hands-on work with development tools, e.g., working according to the change management process in the source code repository and tracing from the architecture modelling tool to requirements and source code.
Finally, Int B was confident that the practice of conducting CIAs guided by a template (cf. Table~\ref{tab:ciatemplate}) ensures comprehensive analyses of high quality.

\begin{table}
\centering
\caption{Overview of RQ2. Black boxes show ``Very useful'', striped boxes ``Moderately useful'', and circles show improvement areas. White boxes mean not mentioned by Int.}
\includegraphics[width=0.375\textwidth]{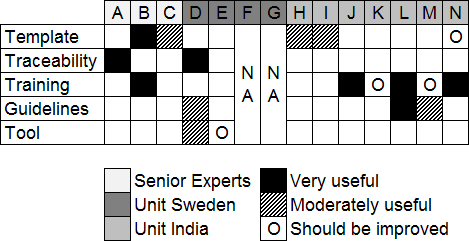}
\label{tab:rq2}
\end{table}

The rest of the Ints expressed neutral opinions of the CIA support.
Three Ints (C, H, I) mention that the CIA template supports the process.
However, some of the Ints are also slightly negative to the template: Int C considers it too cumbersome for his part of the system, Int H no longer needs the template as he already knows its content by heart (``how to answer the questions is really in the back of my mind''), and Int N stresses the importance of letting the template evolve.
Three of the neutral respondents (D, L, M) explain that there are CIA guidelines available on the intranet, providing support on how to fill in the template.
Four Ints (J, L, M, N) appreciate the support they received as new employees, e.g., Int M: ``we received some quality training, I'm equipped with how to answer these questions''
On the other hand, two Ints would have preferred even more training,
as they still rely too much on senior peers when conducting CIA. 
Finally, Ints E and D, both in Unit Sweden, stated that there is a small tool available that allows users to write the CIA report using a form with fixed input fields based on the latest template revision, and it also runs some basic validity checks of user input.

In relation to RQ2, we found several types of support for CIA.
Apart from traceability, discussed in detail in RQ1, engineers report CIA support from 1) training, 2) a CIA template, 3) guidelines, and 4) a prototype tool.
We notice a tendency for the engineers to appreciate the less rigid types of support rather than more formal approaches, i.e., the training and the template is considered more useful than trace links and tool support.
Our interpretation is that the engineers prefer support that allows them flexibility to conduct CIA as they see fit-for-purpose, rather than strict instructions on how to follow trace links, or to be forced into using a specific tool.
On the other hand, when it comes to tool support, the engineers have very limited experience;
Thus, we consider them less likely to bring up tools as as either useful support to CIA or as an important improvement area.
We conclude that the absence of CIA tool support in the case company is in line with what has been reported as state-of-practice~\cite{de_la_vara_industrial_2016}. 


\subsection{Information Seeking Behavior in CIA (RQ3)}
We asked all but one Int about how the engineers seek information in relation to CIA, using a combination of open and closed questions.
The Ints commented on seven approaches to seek information, as listed in Table~\ref{tab:rq3}.

\begin{table}
\centering
\caption{Overview of RQ3. Black boxes show ``Always/Often'', striped boxes ``Sometimes'', and white boxes ``Never/Rarely''.}
\includegraphics[width=0.425\textwidth]{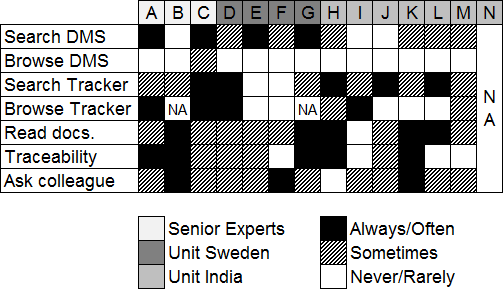}
\label{tab:rq3}
\end{table}

Several information seeking options are standard practice for the Ints: 1) reading documents, 2) asking colleagues, 3) consulting traceability information, 4) searching in the issue tracker, and 5) searching in the DMS. 
All but one Int seek information by \textit{reading documents}, at least sometimes; 
Five Ints do it as part of most CIAs.
Int L explains: ``When conducting CIA, documentation goes side by side. I keep local copies of the documents, and check them whenever I need, for example after working a long time with another component''.
Several Ints use local copies of documents, motivating change alert systems such as suggested by Cleland-Huang \textit{et al.}~\cite{cleland-huang_event-based_2003}. 

Although development in the case company is far from agile, the people perspective is evident.
Several Ints describe CIA as a team effort in which \textit{asking colleagues} is natural, ``especially for old parts with limited documentation or areas with dedicated specialists'' according to Int B.
Only Int H never asks colleagues as part of CIA, motivated by her being the most experienced engineer in the team.

Most Ints use \textit{traceability} when seeking change impact;
Five Ints (A, B, G, H, K) report that it is routinely done (note that Ints A and G also reported traceability as ``very useful'' in RQ1).
Ints report different ways to access trace links (see Section~\ref{sec:RQ1}), and many find them useful.
Int G uses trace links not only to identify change impact, but also to identify colleagues to ask further questions, supporting Hertzum and Pejtersen's finding: engineers' intertwine document and people searching~\cite{hertzum_information-seeking_2000}. 
Int D uses formal traces between requirements and test cases, but does not at all like the T-sheet: ``It is an amoeba, I don't know who is supposed to work with that -- I didn't request it, I don't use it''.
Another reason for not exploiting available traceability in CIA is expressed by Int L: ``I rarely refer to the formal trace links, as I already know them. Also, for older parts of the system the traces are incomplete''.

All but two Ints \textit{search in the issue tracker}, five Ints (C, D, H, J, L) do it always or often.
We discovered two different search approaches;
First, searching for a unique issue ID, i.e., simply extracting information from a known issue.
Second, exploratory searching, i.e., free-text searching, conducted by most Ints. 
However, 4 Ints (D, G, L, M) explicitly complain that the search feature of the issue tracker is inadequate.
Ints D and G complain that only text in the title is indexed, i.e., all descriptive text and also critical information from crash logs are completely missed.
Int M thinks the search queries rarely match the content in the titles.
On the other hand, Int L thinks too many results are returned and that skimming search results in the issue tracker is hard.
On the positive side, Int B likes the filtering options in the tool.

The Ints typically \textit{search in the DMS} as part of CIA, all but three Ints do this at least sometimes and four do it for most cases (A, C, E, G).
We identified the same two distinct types of searching as for the issue tracker: based on document IDs for quick document extraction and exploratory searching.
Several Ints (D, E, G, L, M) frequently try exploratory free-text searches, but they report mixed results. 
Int D estimates that he does exploratory searching when working with older parts of the system, maybe for every 10th CIA, in line with a comment from the safety engineer: ``exploratory searching in the DMS belongs to the past, now you identify document IDs in the architecture model and then just extract them''. 
Int B shares his management perspective, explaining limited need for exploratory searching: ``Due to our static team structure, developers know their respective areas and its documentation. There is no need to search for additional documents''.
Int F expresses critical views on the current DMS searching, sharing that too many results are returned. 
He asks for a concrete information policy, especially regarding document meta-information: ``searching using a tag scheme would be the key to improve our DMS''. 
Also Int L thinks too many results are returned when searching, and states ``after working with Google you are not happy [with the DMS]''.

The two least used ways to seek information addressed in this study both deal with browsing.
Ints share contrasting views on \textit{browsing the issue tracker}: A, C, D, and I do it always or often, while E, F, J, K, and L never or rarely do it.
We note that all Ints that typically browse the issue tracker are senior engineers, suggesting that you need certain experience to fully value the product history.
Motivations for browsing include: it is an efficient way to seek information using the tool's filtering feature (Int C) and it is a useful approach when you are interested in a broad range of old issues (Int H).
Reasons not to browse the issue tracker include: it is tedious (Int K), and as Int L puts it ``I don't browse the issue tracker, I don't like the look of the tool at all -- Just get what you need and get out''.
Only one single Int (C) \textit{browses the DMS}, but he does it only for specific parts of the system.
All others Ints never or rarely browses the DMS, typically explaining that the information structure is highly inaccessible.

We conclude RQ3 by stating that engineers have different information seeking behavior; 
All seeking options are used at least by one Int, and only one Int uses all options.
The identified variation in information seeking was expected, as similar results have been reported in previous work~\cite{hertzum_importance_2002,freund_modeling_2005}.
Our finding highlights the importance of establishing findability in a way that supports different seeking preferences; 
People seek differently, thus the information landscape should offer support accordingly.

Regarding oral communication, we find that it is common information seeking approach in CIA, in line with general expectations on engineers' preferred information seeking~\cite{king_communication_1994};
The fact that dedicated specialists are sometimes consulted in CIA supports Hertzum and Pejtersen's recommendation to not only support searching for documents, but also for people~\cite{hertzum_information-seeking_2000}.
Our results do not support Milewski's findings on cultural differences between India and Europe~\cite{milewski_global_2007};
Ints in Unit India do not ask colleagues more often, and Ints in Unit Sweden do not appear more inclined to read documents.
One possible explanation is that CIA is both a factual and diagnostic task, thus comprehensive information seeking is needed, i.e., using a combination of information sources.

Our recommendation to enable different ways to information seeking is similar to our conclusions for RQ1 and RQ2.
Engineers request flexible support for information seeking in CIA, i.e., instructions on \textit{what} to do, but leaving freedom to choose \textit{how} to do it -- analogous to the nature of both safety standards and requirements specifications.
Trace links are used in CIA at least sometimes by most engineers, but some prefer other ways of seeking change impact.
This supports Cleland-Huang \textit{et al.}'s goal to develop traceability that is fit-for-purpose for an organization~\cite{cleland-huang_software_2014}.
However, our recommendation goes beyond the organizational level: we argue that traceability should be established in ways that let \textit{individual} engineers use it in accordance with their preferred way of working.
How this could be practically achieved needs further research, but we envision maintenance of multifaceted traceability information that could be accessed in different ways for different purposes.
We argue that making access to trace links task-based is a promising approach, in the same way as it has been beneficial in information retrieval~\cite{freund_modeling_2005,grzywaczewski_task-specific_2012}.


\subsection{Satisfaction Assessment in CIA (RQ4)} 
When preparing the interviews, we selected (when applicable) two previously completed CIA reports for each Int representative of: considerable impact on non-code artifacts (Type~1), and no impact at all on non-code artifacts (Type~2).
We asked the Ints to describe how they conducted the CIAs, as a preparation for a question on satisfaction criteria;
For the two specific CIA reports, we asked 10 out of 14 Ints to describe how they assured that they had reported all relevant change impact in Type 1 and/or how they determined that there was indeed no non-code change impact in Type~2.

Regarding both completeness of CIA reports of Type 1, and the validity of reporting no impact in CIA reports of Type 2, the interviewees rely much on experience. 
Three Ints (B, C, F) consider it a a difficult problem.
For Type 1, Ints J and M claim that they already know the related documentation well prior to the CIA.
Int J remembers the content of DDs from the formal document reviews in earlier phases.
Int M explains that he knows the contents of the test case specifications from repeatedly running them: ``Mostly I know which test cases are there, I work a lot with them. For each delivery we run half of the functional test cases''.
Int E shares that when he feels done with the CIA, he concludes with exploratory searching for keywords to check if any new requirements have been added to the involved documents.

Most Ints shared their views on reporting no non-code impact in Type 2 CIAs.
Again several Ints emphasize experience (A, B, E), e.g., Int B: ``It's probably very much based on experience. You have the whole spectrum from minor bug fixes to complex changes, that's the problem. You must have completed many CIAs to know when answering that `no impact' is reasonable.'' Int B also adds ``Our developers are skilled, if they say no impact I trust them''.
Other interviews (C, H, M) confirm that `no impact' is easy to conclude for minor bug fixes, including variable initialization and renaming and adding null pointer checks.

Int A explains that CIAs are mainly conducted by developers with expertise on the component, typically they have both implemented the features and authored the accompanying documentation.
However, Ints A and D says that for old parts of the system there is limited documentation, in particular no DDs.
Ints B and D both stress that DDs should not cover the finest level of implementation details, Int D also says that you conclude `no impact' ``when you feel that there are no documents covering the area you are about to change. That is, you don't want to author a new document or you don't think the documents should cover this level of detail.''

Ints F and G share more hands-on approaches to assure that `no impact' is correct.
Int F searches the architecture model, and also the related documents;
He explains that he tries to trace all the way to where the feature should be described, and if he finds nothing he also asks colleagues.
Nonetheless, he adds ``Either there was nothing to find, or I didn't find it'' when motivating his Type 2 CIA.
Int G also frequently asks colleagues to conclude a Type 2 CIA, but for the specific CIA he quickly determined that the change was limited to a simulated environment tailored for a specific customer: ``This change was mostly politics [sic] to make the simulator more like the real product''.

Concluding that a CIA report is complete is analogous to satisfaction assessment as described by Holbrook \textit{et al.}~\cite{holbrook_study_2013}, i.e., checking that a set of trace links encompasses every relationship that should be present.
They state that satisfaction assessment is mainly done manually, which is corroborated by our study;
We show that engineers mainly rely on experience when doing satisfaction assessment, a finding that opens up opportunities for complementary tool support.
Furthermore, our study shows that satisfaction assessment is not only important for entire traceability matrices, but also for smaller tracing efforts such as CIA.
Also Niu \textit{et al.}'s work on information seeking when vetting traceability matrices is related to this RQ~\cite{niu_departures_2013}, but their work focused on confirming candidate trace links rather missing links.

\section{Threats to Validity}
\label{sec:threats}

This section discusses threats to validity in line with Runeson \textit{et al.} \cite{runeson_case_2012}, with focus on construct, internal, external and reliability issues.
\textit{Construct validity} reflects how well the phenomenon under study is captured. The main researcher in the study has earlier work experience in one of the studied sites, and spent more than a month in the other site for the study to make sure that the constructs were well understood (prolonged involvement). Regarding the constructs under study, our scope of focusing on non-code artifacts may be too restrictive. However, interviewees were able to add additional aspects in the open interviews, although we did not specifically ask for code aspects. Further, the list of seeking options might be too limited, but again, there was the option to add more. The results indicate that the variations are large within the options we proposed, and we do not conclude anything specific about the seeking patterns. Cultural differences might also lead to bias to answers in the two units. However, based on the analysis, we do not see any of the expected cultural differences influence the outcome. Probably, the company culture is as strong as the local culture.

\textit{Internal validity} relates to causal relationships. We do not claim any such relations, only hypothesize correlations between factors. For example, in different patterns between sites, we add complexity of tasks performed as a hypothetical difference. This is also mitigated by the researcher's prolonged involvement. When it comes to \textit{external validity}, this is related to generalization of the findings outside the studied setting. We do not aim for generalizability beyond the domain of safety-critical systems, since the demands for CIA is regulated for this domain. However, since the variation was large within one company, even within a regulated context, we are confident that variations between companies may be similar. Validation studies are always appreciated, for example including other roles or companies, but we consider this be further work.

The \textit{reliability} of the study is related to the dependence on specific researchers. Actions taken in this case to mitigate threats to reliability include to apply established research practices to conduct interviews based on a scheme, keep audit trail of data, including recording, transcription, independent analysis, traceability of conclusions etc. Specifically, as a qualitative study focuses on understanding a phenomenon -- in our case information seeking behavior in CIA -- other researchers might find interest in other details, but the core findings are on a quite high level of detail. 






\section{Conclusion}
\label{sec:conc}

Supporting findability in Software Engineering (SE) projects is essential.
The most useful tool support is often task dependent, thus researchers must first understand engineers' information seeking behavior in specific tasks.
We conducted a case study with two units of analysis on how engineers seek information as part of Change Impact Analysis (CIA), known to be a knowledge-intensive task involving a cognitive process.

We showed that \textit{engineers use several different types of trace links in CIA} (RQ1).
However, \textit{all engineers do not consider traceability particularly useful} when conducting CIA, in contrast to the generally positive claims in previous studies~\cite{de_lucia_traceability_2008,li_requirement-centric_2008,cuddeback_automated_2010};
Some engineers are indifferent, i.e., they do not recognize a need for trace links in their part of the information landscape.
We conclude that the existence of trace links is not enough to provide support for CIA -- companies should make an effort to also make traceability accessible.
Although more research is needed, we hypothesize that engineers often consider trace links to be more of a burden than a help. 

Apart from traceability, we identified a handful of other approaches to support CIA in the case company (RQ2): \textit{a dedicated CIA template, formal training, written guidelines, and a prototype tool}.
Thus, in line with previous work~\cite{de_la_vara_industrial_2016}, we confirm that CIA is mainly a manual activity in industry. 
Moreover, we notice a \textit{tendency for the engineers to prefer less rigid types of support more than formal approaches}. 
We believe that engineers prefer support that leaves them flexibility to conduct CIA as they see fit-for-purpose, rather than a regulated process on how to trace impact, or to be forced into using a specific tool.

We showed that \textit{engineers have different information seeking behavior in CIA} (RQ3).
The varied individual preferences stress the \textit{importance of supporting different seeking approaches}, such as both searching and browsing in databases, and following trace links.
All but one interviewee seek information by asking colleagues, resonating with previous research stating that \textit{engineers are inclined to oral communication}~\cite{king_communication_1994} and \textit{intertwine looking for documents and people}~\cite{hertzum_information-seeking_2000}.

We identified two opportunities to improve findability in the case company.
First, several interviewees expressed clearly negative views on the old-fashioned issue tracker in use;
Improving it could be an important investment, e.g., by integrating full-text indexing of issue reports~\cite{borg_replicated_2014}.
Second, only one single interviewee browses the document management system, the others complain about its information structure.
Thus, one basic way to seek information, i.e., browsing, appears to be mostly unused in the case company.
Improving the browsing user experience, combined with developing an information policy in close collaboration with the engineers intended to use the database, could improve findability -- not only for CIA, but for SE in general.

Finally, we found that engineers' CIA satisfaction assessment, i.e., concluding that all relevant change impact has been identified, primarily is \textit{based on experience and gut feeling} (RQ4).
This finding is in line with previous work on satisfaction assessment of traceability matrices, reported to be mostly a manual activity~\cite{holbrook_study_2013}.
As development that mandate CIA is typically safety-critical, gut feeling appears to be an insufficient approach.
We have previously proposed tool support for CIA satisfaction assessment~\cite{borg_supporting_2016}, but future work is needed to explore how such tools can be introduced without reducing engineers' flexibility as to how they conduct CIA.
If anything, our study points out individual variations of how engineers prefer to work in CIA;
Future CIA support would benefit from enabling different ways of working, rather than shoehorning engineers into a single impact seeking process. 

\section*{Acknowledgment}
The authors would like to thank all participants in the case study.
The work is partially supported by a research grant for the ORION project (reference number 20140218) from The Knowledge Foundation in Sweden, and the Industrial Excellence Center EASE - Embedded Applications Software Engineering\footnote{http://ease.cs.lth.se}. 



\bibliographystyle{IEEEtran}
\bibliography{phd}
%



\end{document}